\documentclass[twocolumn,tighten,times]{aastex631}

\usepackage{amsmath}
\usepackage[T1]{fontenc}
\usepackage[utf8]{inputenc}
\usepackage{natbib,hyperref}
\graphicspath{{./}{figures/}}

\usepackage{subfigure}

\shorttitle{Electron-ion equipartition in relativistic shocks}
\shortauthors{Vanthieghem et al.}

\begin{document}

\title{Origin of intense electron heating in relativistic blast waves}

\correspondingauthor{Arno Vanthieghem}
\email{vanthieghem@astro.princeton.edu}

\author[0000-0002-3643-9205]{Arno Vanthieghem}
\affil{International Research Collaboration Center, National Institutes of Natural Sciences, Tokyo 105-0001, Japan}
\affil{Department of Astrophysical Sciences, Princeton University, Princeton, NJ 08544, USA}
\affil{High Energy Density Science Division, SLAC National Accelerator Laboratory, Menlo Park, California 94025, USA}
\affiliation{Sorbonne Universit\'e, Institut Lagrange de Paris (ILP), 98 bis boulevard Arago, F-75014 Paris, France}
\affiliation{Institut d'Astrophysique de Paris, CNRS -- Sorbonne Universit\'e, 98 bis boulevard Arago, F-75014 Paris, France}

\author[0000-0002-2395-7812]{Martin Lemoine}
\affiliation{Institut d'Astrophysique de Paris, CNRS -- Sorbonne Universit\'e, 98 bis boulevard Arago, F-75014 Paris, France}

\author[0000-0003-0116-5248]{Laurent Gremillet}
\affiliation{CEA, DAM, DIF, F-91297 Arpajon, France}
\affiliation{Universit\'e Paris-Saclay, CEA, LMCE, 91680 Bruy\`eres-le-Ch\^atel, France}



\begin{abstract}
The modeling of gamma-ray burst afterglow emission bears witness to strong electron heating in the precursor of Weibel-mediated, relativistic collisionless shock waves propagating in unmagnetized electron-ion plasmas. In this Letter, we propose a theoretical model, which describes electron heating  via a Joule-like process caused by pitch-angle scattering in the decelerating, self-induced microturbulence and the coherent charge-separation field induced by the difference in inertia between electrons and ions.
The emergence of this electric field across the precursor of electron-ion shocks is confirmed by large-scale particle-in-cell (PIC) simulations. 
Integrating the model using a Monte Carlo-Poisson method, we compare the main observables to the PIC simulations to conclude that the above mechanism can indeed account for the bulk of electron heating.
\end{abstract}



\section{Introduction} \label{sec:introduction}

\begin{figure*}[t]
	\begin{center}
		\includegraphics[width=0.95\textwidth]{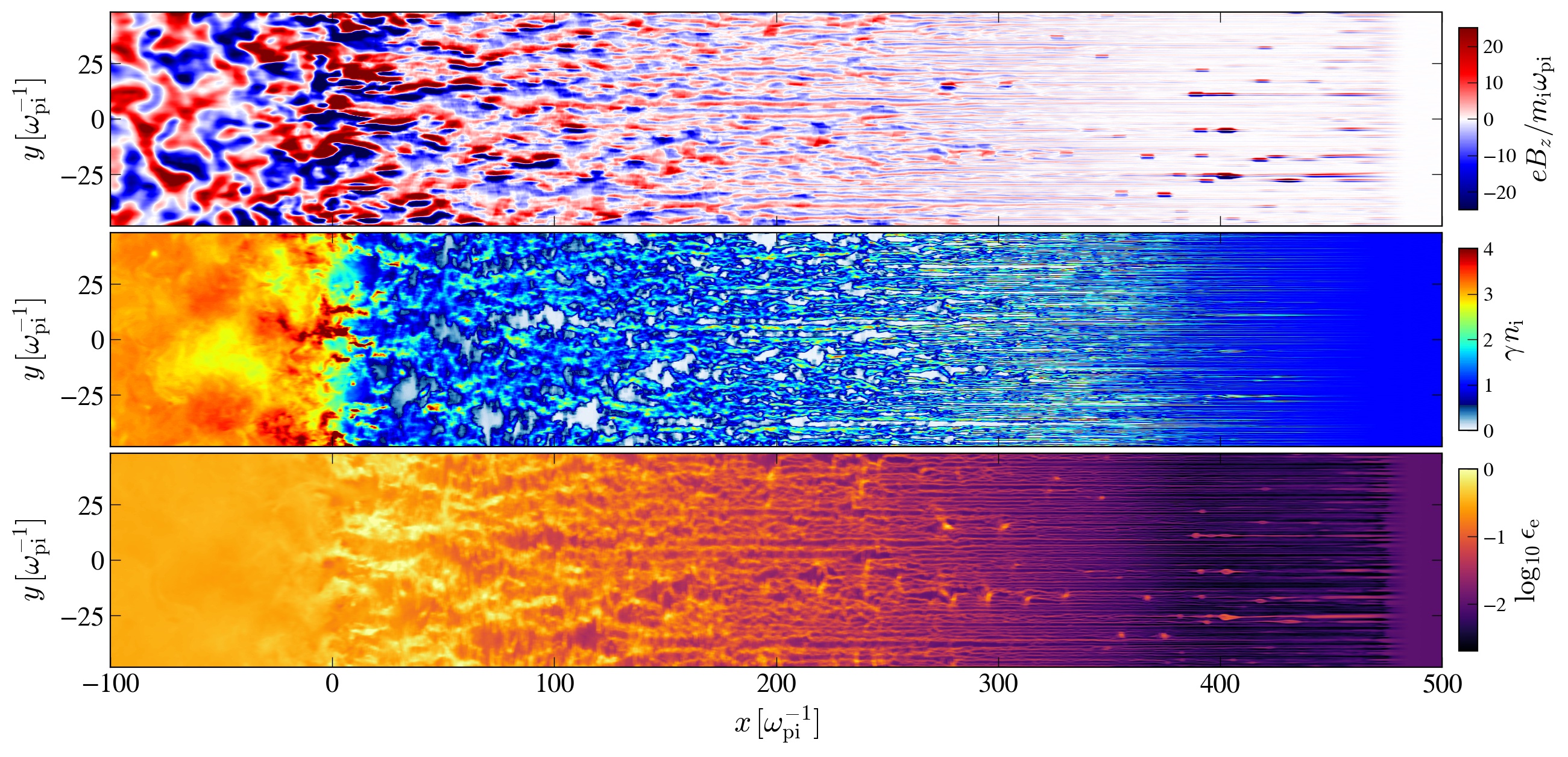}
		\caption{Closeup of the precursor of an unmagnetized  electron-ion collisionless shock in the PIC simulation frame. The upstream plasma is injected from the right-hand side with a Lorentz factor $\gamma_{\rm \infty}=100$, proper temperatures $T_{\rm e}=T_{\rm i}=0.01 m_{\rm e}$, and mass ratio $r=m_i/m_e=100$. The top (resp. middle and bottom) panel shows the magnetic field (resp. ion density and electron energy fraction) profile at time $t=870\,\omega_{\rm pi}^{-1}$. The shock front, located at $x=0$, separates the unshocked plasma ($x>0$) from  the shocked plasma ($x<0$).
} 
		\label{fig:prof_ga100}
	\end{center}
\end{figure*}

The dissipation of the bulk energy of astrophysical outflows into nonthermal distributions of accelerated particles appears both generic and multifarious in the high-energy Universe. Collisionless shock waves account for such dissipation in varied environments, from our own Solar system to extreme relativistic events such as gamma-ray bursts (GRB). At the external boundary of the GRB jet, a shock front indeed sweeps the external medium at a velocity close to that of light, and eventually radiates part of the blast energy into multiwavelength spectra through synchrotron-self-Compton radiation of shock-accelerated electrons~\citep[][and references therein]{Kumar_2015a}. These
emissions have by now been detected up to TeV energies~\citep{2019Natur.575..455M,2019Natur.575..464A} and, in one outstanding case, they have formed the electromagnetic counterpart of a gravitational wave event~\citep{Abbott_2017b}, thereby laying a foundation stone of multi-messenger astrophysics. 

Accordingly, this calls for an improved understanding of the physics of relativistic collisionless shock waves, the intricacy of which lies in the subtle intertwining of the self-generated electromagnetic microturbulence that mediates the dissipative dynamics of the dilute plasma, with the nonthermal beam of accelerated particles that drives the microturbulence in which particles scatter and possibly radiate~\citep[][for a review]{Marcowith_2016}. One striking feature of weakly magnetized, relativistic collisionless shocks is their ability to convert a fraction as large as $10-30\,\%$ of the shock dissipated energy into nonthermal high-energy electrons. This finding, confirmed in particle-in-cell (PIC) numerical experiments~\citep{Spitkovsky_2008, Martins_2009, Haugbolle_2011, Sironi_2013}, nicely accounts for the large inferred radiative efficiency of GRB afterglows~\citep{2001ApJ...547..922F}, and is therefore of prime interest for astrophysical phenomenology. For comparison, this fraction falls by one to two orders of magnitude in subrelativistic shock waves, such as those formed in supernovae remnants~\citep{2002A&A...396..649V}; and, should electrons verify the shock crossing conditions independently of ions, it would equal the electron-to-ion mass ratio $m_e/m_i$. 

As viewed from the reference frame of the shock front, the incoming plasma energy is mostly carried by the ions, but the electrons can sap energy from this reservoir through their interaction with the electric fields generated in the shock precursor and transition layer. Yet the
underlying mechanism  remains widely debated: the scenarios proposed so far rely on the inductive electric field associated with the growth of the primary current filamentation instability (CFI)~\citep{Gedalin_2012, Kumar_2015} or of a secondary kink-type instability~\citep{Milosavljevic_2006a}, as well as on the longitudinal~\citep{Gedalin_2008, Plotnikov_2013, Kumar_2015} and/or transverse~\citep{Plotnikov_2013} electrostatic components that accompany the CFI and its oblique variants~\citep{Bret_2008}. 

The objective of this Letter is to clarify the physics of electron heating in relativistic electron-ion collisionless shocks. 
To do so, we have performed large-scale particle-in-cell (PIC) simulations of relativistic, unmagnetized electron-ion shock waves (\S\ref{sec:PIC}). In parallel, we have developed a theoretical model, which generalizes a previous model for 
\emph{pair} shocks~\citep{Lemoine_2019_PRL, Pelletier_2019, Lemoine_2019_II, Lemoine_2019_III}, to 
describe the deceleration and heating dynamics of the background plasma electrons and ions (\S\ref{sec:modsim}). The numerical integration of this model, through a Monte Carlo-Poisson technique, gives results that compare satisfactorily well with the PIC simulations. In this picture, electron heating can be depicted as a collisionless Joule process in a large-scale, longitudinal electric field due to charge separation in the background plasma, itself entailed by the differential scattering strength of electrons and ions in the Weibel microturbulence.
Our results are summarized and discussed in \S\ref{sec:discussion}. Throughout we use units in which $k_B=c=1$.

\section{Kinetic simulations} \label{sec:PIC}

\begin{figure*}[ht]
	\begin{center}
		\includegraphics[width=1.\textwidth]{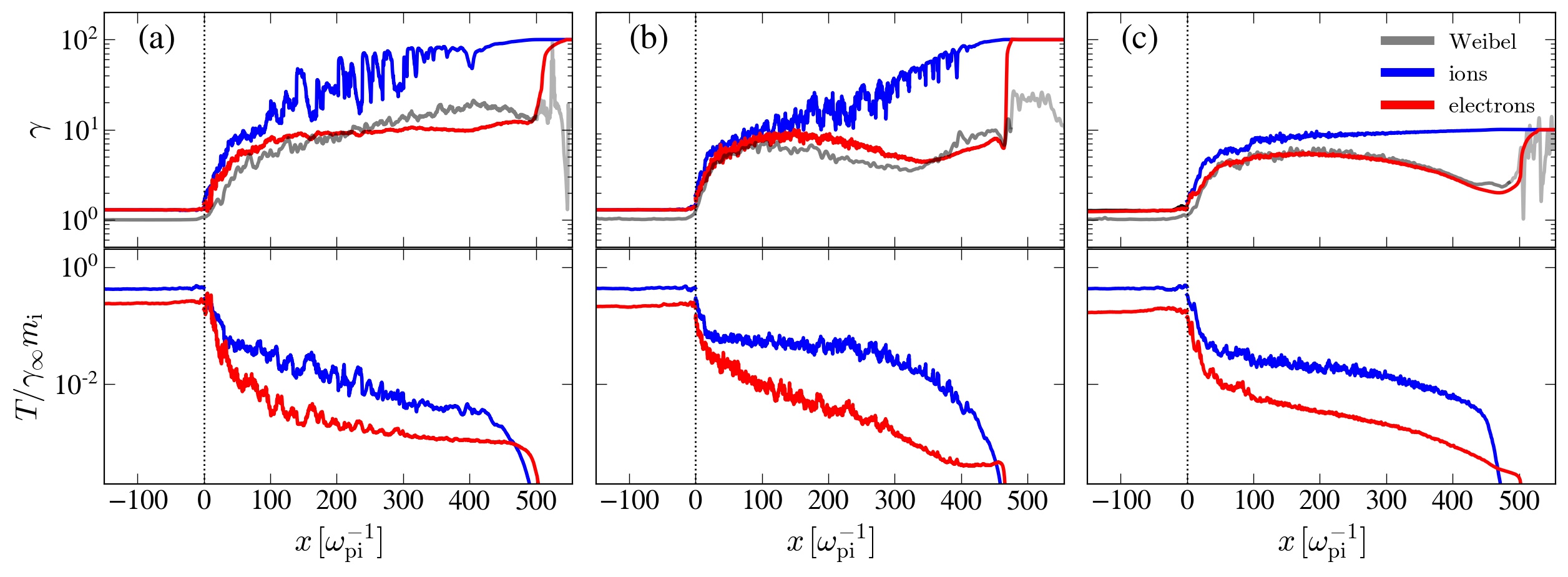}
		\caption{Drift Lorentz factors (top) and temperatures (bottom) for ions (blue) and electrons (red) of the background plasma, extracted from our three reference PIC simulations with parameters: (a) $\gamma_\infty =100$ and $r=25$; (b) $\gamma_\infty =100$ and $r=100$; (c) $\gamma_\infty =10$  and $r=100$. In the top panels, the solid gray lines  plot the Lorentz factor of the Weibel frame as extracted from each simulation (see text).
} 
		\label{fig:PIC_profs}
	\end{center}
\end{figure*}

Our kinetic simulations are performed in 2D3V (2D in physical space, 3D in momentum space) geometry using the finite-difference time-domain, relativistic PIC \textsc{calder} code~\citep{Lefebvre_2003}, which has been extensively used in the relativistic regime~\citep[\emph{e.g.},][]{Vanthieghem_2018}, where it has been shown to properly expunge the salient relativistic beam-grid instabilities by means of the Cole-Karkkainen electromagnetic solver~\citep{Karkkainen_2006} coupled with the Godfrey-Vay filtering method~\citep{Godfrey_2014}. 
The plasma is injected from the right-hand side of the box with a negative relativistic velocity $\beta_\infty$ (corresponding Lorentz factor $\gamma_\infty$). To further reduce computational time, the domain is progressively lengthened using a moving injector, keeping electromagnetic waves and reflected particles away from the right boundary. On the left-hand side, we impose perfect conductor conditions for the fields and specularly reflecting conditions for the particles. We present here the results of three large-scale simulations with various mass ratios $r = m_{\rm i}/m_{\rm e} = (25,100,100)$ and Lorentz factors $\gamma_\infty = (100,100,10)$; in each, the plasma is initialized with low proper temperatures $T_{\rm i}= T_{\rm e} = 10^{-2} m_{\rm e}$. The simulation frame corresponds to the downstream rest frame ($_{\vert \rm d}$ subscript), hence $\gamma_\infty$ represents the relative Lorentz factor between upstream and downstream. Correspondingly, the shock Lorentz factor, measured relative to upstream is $\sqrt{3}\gamma_\infty$ for a relativistic shock in 2D3V. The relativistic electron skin depth $c/\omega_{\rm pe}$ of the unshocked plasma is resolved with a mesh size $\Delta x = 0.08 c/\omega_{\rm pe}$ and $\Delta t = 0.99 \Delta x$.
Our simulations use 10 particles per cell and per species, and cover a duration $L_t \simeq 10^3\,\omega_{\rm pi}^{-1}$ ($\omega_{\rm pi}$ the upstream ion plasma frequency), corresponding to $\simeq 120\,000$ iterations, at which time the numerical domain comprises $120\,000\times 12\,000$ cells. To enable proper comparison with the model, we distinguish the background plasma particles from the nonthermal population as in \cite{Lemoine_2019_PRL}: the background plasma is defined as the set of particles with negative velocity ($\beta_x \leq 0$), which never experienced any turnaround in the microturbulence.

All simulations exhibit the characteristic filamentary pattern of the CFI, which is seen to grow in amplitude from the far upstream towards the shock front, as a result of the interpenetration of the beam of accelerated particles with the background plasma (Fig.~\ref{fig:prof_ga100}, top and middle panels). Across the precursor, the background plasma electrons are continuously energized to the point of reaching quasi-equipartition when crossing the shock front -- \emph{i.e.} the energy partitions $\epsilon_\alpha = e_\alpha/(e_e+e_i) \sim \mathcal O(1)$ where $e_\alpha$ represents the energy density of species $\alpha$ (Fig.~\ref{fig:prof_ga100}, bottom panel).

The transversely averaged drift Lorentz factor and temperature profiles for the background plasma species (Fig.~\ref{fig:PIC_profs}) show that, in each case, the electrons are abruptly slowed down upon entering the shock precursor, and progressively heated up to $T_{\rm e} \simeq 0.5 T_{\rm i}$ in the downstream, where the ion temperature matches the (2D relativistic) hydrodynamic jump conditions: $T_{\rm i}/m_{\rm i} = \gamma_\infty/2$.
For $\gamma_\infty = 100$ and $r = (25,100)$, the ion Lorentz factor shows a regular decrease until reaching  the (subshock) transition layer ($x\simeq 50 c/\omega_{\rm pi}$) where the $\gamma_{\rm i}$ and $\gamma_{\rm e}$ curves converge with each other and rapidly drop to $\sim 1$. For $\gamma_\infty = 10$ and $r = 100$, the ions undergo only weak deceleration over most of the precursor, and are hence mainly stopped in the transition layer.

\begin{figure}[t]
	\begin{center}
		\includegraphics[width=1.\columnwidth]{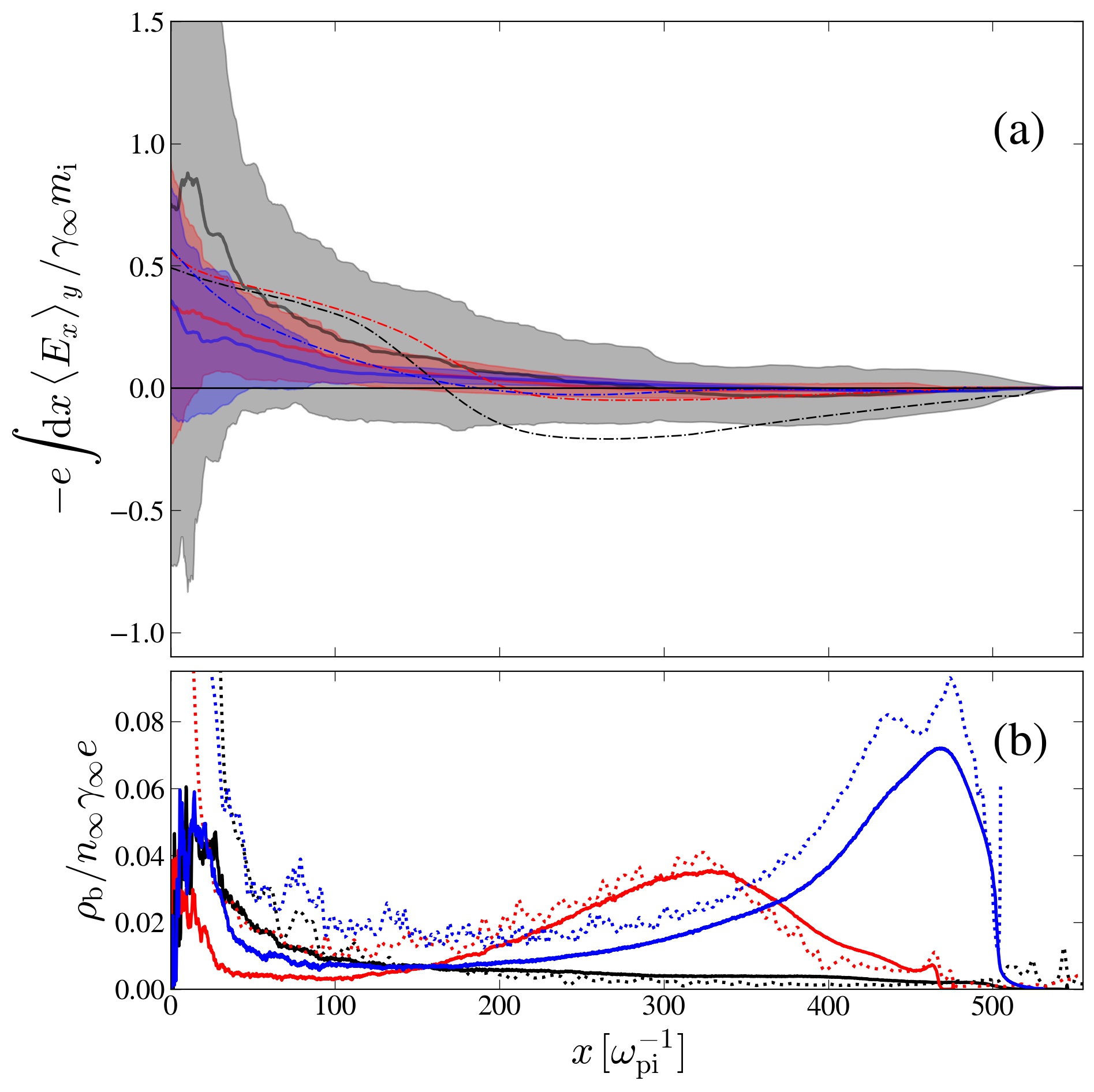}
		\caption{
		(a) Estimate of the coherent mean electric potential $ -\int \langle E_x \rangle_y\,{\rm d}x$  weighted by the electron density (solid curves) and in the range of $\pm 1 \sigma$ variation (shaded area).  The different colors respectively correspond to the three PIC simulations with  $(r,\gamma_\infty)= (100,10)$ (blue), $(25, 100)$ (black), and $(100,100)$ (red). The dot-dashed lines correspond to the electrostatic potential reconstructed from the Monte Carlo-Poisson integration of our theoretical model. (b) Charge density of the beam component as measured in the PIC simulations with the same color code as in panel (a). The suprathermal beam is defined as the ensemble of particles moving with longitudinal velocity $\beta_x > 0$. The dotted lines show the relationship between the (apparent) beam charge density $\rho_{\rm b}$ and $\gamma_{\rm w}$, as extracted from Eq.~\eqref{eq:beam_nonneutral}, namely, $ \rho_{\rm b} \propto \, \gamma_\infty n_\infty\,e/\gamma_{\rm w}^2$, which matches well the observed profile.
}
		\label{fig:rho_phi}
	\end{center}
\end{figure}

A key observation, specific to the unmagnetized regime, is the existence of a particular reference frame -- the Weibel frame ($\mathcal{R}_{\rm w}$) -- in which the microturbulence is essentially magnetostatic, because generated by the CFI~\citep{Pelletier_2019}. In the simulation frame, this Weibel frame strongly decelerates from a large four-velocity $\left\vert u_{\rm w}\right\vert\sim\gamma_\infty$ at the tip of the precursor to low subrelativistic $\left\vert u_{\rm w}\right\vert\sim 0.5$ values at the shock transition.
The dynamics of that Weibel frame can be extracted from the PIC simulations, in particular its $3-$velocity $\beta_{\rm w}(x) = -\langle\delta E_y^2\rangle_y^{1/2}/\langle \delta B_z^2\rangle_y^{1/2}$ (sign fixed by our conventions).

The $x-$profile of the corresponding Lorentz factor $\gamma_{\rm w}$ is plotted (in gray) in the top panels of Fig~\ref{fig:PIC_profs}. In each simulation case, the Weibel and electron Lorentz factors exhibit similar dynamics: this corroborates that the background plasma electrons are strongly coupled to the turbulence throughout the precursor, unlike the ions whose larger inertia lets them stream well ahead. This effect will be shown to be mainly responsible for the generation of a coherent, longitudinal electric field which will eventually lead to intense electron heating.

Our PIC simulations confirm the generation of a coherent longitudinal electric field $\langle E_x \rangle_y$ across the precursor. Figure~\ref{fig:rho_phi}(a) plots the spatial profile of the associated electric potential, $\langle \phi \rangle_y = -\int \langle E_x \rangle_y\,{\rm d}x$, in the three simulation cases\footnote{ To extract the electric potential as seen by the electrons, we weight the electric contribution at each point by the electron density of the background plasma. This compensates for the slight transverse heating inhomogeneities visible in Fig.~\ref{fig:prof_ga100}. The potentials calculated without this weighting techniques take lower values at the shock, but the coherent longitudinal electric field remains.}. 
The potential is shown in solid lines, and the shaded areas indicate the $\pm1\sigma$ variations of $-\int E_x\,{\rm d}x$ (again, weighted by the electron density, but without the average over $y$) in the transverse direction. Importantly, this potential exhibits a net mean value, whose  magnitude reaches $\langle\phi\rangle_y \simeq 0.2 - 0.5\, \gamma_\infty m_i /e$. It is sufficient, in sign and in order of magnitude, to explain electron heating up to near equipartition. In the following, we develop a theoretical model to interpret the main features of those PIC simulations, {\it i.e.} the deceleration and heating of electrons and ions, the origin of this electric field and its role regarding electron heating.

We interpret this coherent electric field as resulting from two causes (see also below):
(i) to a large extent, the charge separation between electrons and ions in the background plasma owing to their different degree of coupling to the turbulence; (ii) to a lesser extent, at larger distances from the shock, the (apparent) net charge density, $\rho_{\rm b}$, carried by the beam of suprathermal particles. The latter charge imbalance is weak in the simulation with $r=25$ [red line in Fig.~\ref{fig:rho_phi}(b)], but significant for $r=100$ (black and blue lines). In the simulations, it originates primarily from the initial stage of shock formation but, in actual situations, it can be sustained by differential injection of electrons and ions at the shock front. We thus take it into account and weigh its relevance against charge separation between the background electrons and ions by comparing the three simulations. 

One can expect other contributions to $\langle E_x\rangle_y$, in particular an inductive electric field associated with the CFI growth, or an electrostatic field linked to the broadband nature of the CFI. Qualitatively, those are expected to give contributions of opposite signs in current filaments of opposite polarity, thus translating into a small net value when directly averaged over the transverse direction. The field strength measured without weighting by the electron density supports the idea that it rather originates from charge separation, as discussed above. The successful comparison of our model to the PIC simulations will further corroborate this idea.

The above two sources of charge imbalance act differently on the background electrons and  on the electric potential: the positively charged suprathermal beam tends to repel the electrons as it moves away from the shock, while the background plasma ions moving in the other direction tend to drag them toward the shock. 
Their zones of influence can be read off the bottom panel of Fig.~\ref{fig:rho_phi}. 
The charge density profile impacts the bulk dynamics of the background electrons,
because of approximate charge and current neutrality at every point (as observed in the simulations). To see this, consider three populations: the electrons and ions of the thermal background and a beam of suprathermal particles carrying charge density $\rho_{\rm b}$. To a good approximation, the particle current density of the background ions is conserved along the precursor 
(verified in the simulations),  $n_{\rm i} u_{\rm i} = n_{\rm \infty} u_{\rm \infty}$. Then, imposing overall charge and current (quasi-)neutrality leads to 
\begin{equation}\label{eq:beam_nonneutral}
\gamma_{\rm e}^2 = \frac{\gamma_{\rm i}^2}{\left[ 1+ 2  \gamma_{\rm i}^2\left( 1 + \beta_{\rm b} \right) \displaystyle{\frac{\rho_{\rm b}}{\gamma_\infty n_\infty e}}  \right]}\,,
\end{equation}
hence $\gamma_{\rm e}^{-2}\propto \rho_{\rm b}$ in the far precursor where $\rho_{\rm b}/ \gamma_\infty n_\infty e\gtrsim\,\gamma_\infty^{-2}$. 
Since $\gamma_{\rm e} \sim \gamma_{\rm w}$ (see Fig.~\ref{fig:PIC_profs}), the above relation implies $\rho_b \propto \gamma_{\rm w}^{-2}$. Figure~\ref{fig:rho_phi}(b) demonstrates that this scaling law is nicely verified in our simulations.  

The drift speed of $\mathcal{R}_{\rm w}$ in which the background electrons relax through scattering, depends non-trivially on the physical characteristics of the beam and the background plasma~\citep{Pelletier_2019}; in the presence of a net charge, this relation becomes even more complex, because of charge compensation, as discussed above. The drift velocity of $\mathcal{R}_{\rm w}$ is predicted to be smaller (in magnitude) than that of the background plasma; this notably explains the sharp deceleration of background electrons at the tip of the precursor\footnote{Note that Figure~\ref{fig:PIC_profs} displays the drift Lorentz factors and temperatures. When plotting the average Lorentz factor, one recovers the behavior shown in~\citep{Spitkovsky_2008, Haugbolle_2011}.}, where they first penetrate this microturbulence. 

\section{Physics of electron heating}
\label{sec:modsim}

We model the physics of electron heating in the microturbulence of the shock precursor as follows. 
In the decelerating Weibel frame $\mathcal R_{\rm w}$, particles are subject to an  effective gravity, to the coherent electric field $\langle E_x\rangle_y$ and to angular scattering off the microturbulence. This combination gives rise to efficient heating through a collisionless Joule process, in which the gravity and the electric field serve as driving forces along $x$, while scattering redistributes the energy gained or lost in the transverse directions. The heating induced by the effective gravity can be understood, in the simulation frame, as due to the perpendicular motional electric fields carried by the filamentary structures~\citep{Lemoine_2019_II}. We show below that this contribution accounts for the bulk of ion heating, while the longitudinal coherent electric field will be responsible for most of electron heating\footnote{The transverse electromagnetic fields, as extracted from our simulations to define $\mathcal{R}_{\rm w}$, encompass perpendicular electric fields of any nature. While their contribution is subdominant relative to Weibel modes, the perpendicular electric fields associated with oblique instabilities and their effect are thus, to some extent, captured in our analysis.}.

Of course, the electric field does not only heat the electrons, it also slows down the ions and accelerates the electrons toward the shock. Therefore, not all of the magnitude of $e\phi$ at the shock is converted into electron heating. To understand the detailed contribution of the electric field to electron heating, we extend the Fokker-Planck description of \cite{Lemoine_2019_II} by incorporating this longitudinal component. To first order in the relative velocity $\vert\beta_{\rm e\vert w}\vert$ between the background electrons and $\mathcal R_{\rm w}$, we find that this Joule heating can be characterized by the momentum diffusion coefficient (for ultrarelativistic electrons)
\begin{equation}
    D_{pp}\,=\,\frac{\beta_{\rm w\vert s}^2 p_{\vert\rm w}^2}{3\nu_{\rm e\vert w}} \left[\frac{{\rm d}u_{\rm w\vert s}}{{\rm d}x_{\vert\rm s}}\left(1 - \frac{ 1}{3\beta_{\rm w\vert s}^2 }  \right)  - \frac{q E_x}{\beta_{\rm w\vert s} p_{\vert\rm w}}\right]^2
    \label{eq:Dpp}
\end{equation}
with the following notations: quantities indexed with $_{\vert\rm s}$ (resp. $_{\vert\rm w}$) are understood to be defined in the rest frame of the shock (resp. $\mathcal R_{\rm w}$); $p_{\vert\rm w}$ represents the particle momentum in $\mathcal R_{\rm w}$.  The first term in the brackets is the inertial term associated with the deceleration of $\mathcal R_{\rm w}$ (effective gravity), while the second represents the contribution of the coherent longitudinal electric field to the driving force. Such a Fokker-Planck analysis is well-suited to the electrons here, because their drift velocity in the Weibel frame $\vert\beta_{\rm e\vert w}\vert\ll1$, see \S\ref{sec:PIC}. 

Interestingly, Eq.~(\ref{eq:Dpp}) indicates that the sign of the electric field does not matter much for what regards heating through the Joule effect. This agrees, at least qualitatively, with the observation that electrons are systematically heated over the precursor, even in regions in which $\langle E_x\rangle_y$ is negative because of the influence of the nonneutral particle beam (Fig.~\ref{fig:PIC_profs}).

\begin{figure}[htb]
	\begin{center}
		\includegraphics[width=1.\columnwidth]{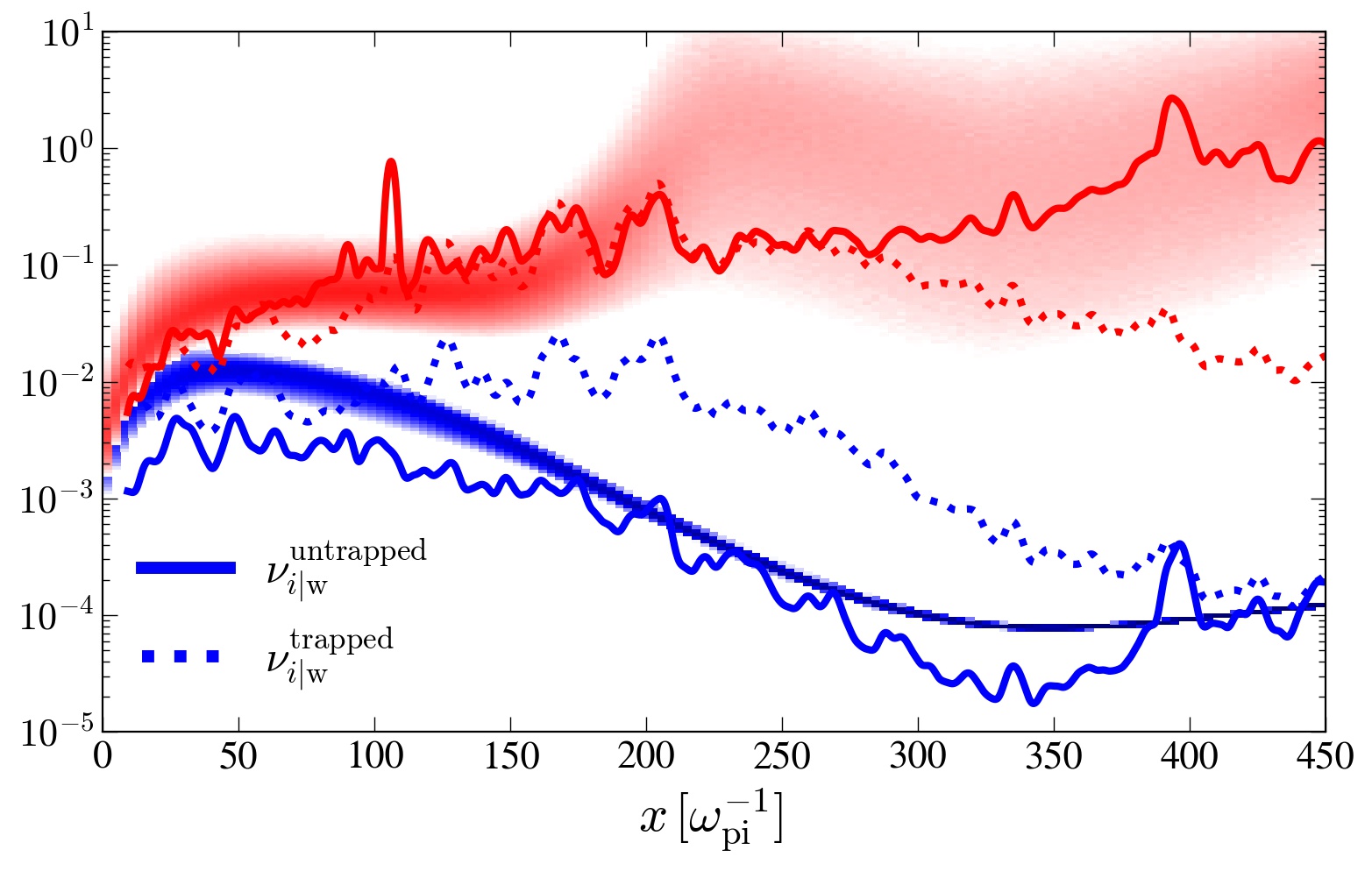}
		\caption{In shaded bands, distributions of the electron (red) and ion (blue) scattering frequencies across the shock precursor as extracted from the MC-Poisson solution to the transport equation using the continuous formula \eqref{eq:MC_nu} with $(r,\gamma_\infty) =(100,100)$. For comparison are overlaid the scattering frequencies in the trapped (dotted lines) and untrapped (solid lines) limits as extracted from the corresponding PIC simulation using Eqs.~(34) and (35) of ~\cite{Lemoine_2019_II}. 
		}
		\label{fig:nu}
	\end{center}
\end{figure}

To validate our theoretical model against PIC simulations, we numerically integrate it using a Monte Carlo-Poisson (MCP) method, which solves the transport equation, including the requisite physical ingredients of our theoretical model, while discarding other kinetic effects. 
We then extract the main observables, namely the profiles of the four-velocity and temperature of each species along the shock normal, and compare them to those seen in the PIC simulation.

Specifically, the model describes the kinematics of the background plasma particles (in $\mathcal R_{\rm w}$) through the (discretized) stochastic equations
\begin{align}
    {\rm \Delta} \mu_{\alpha|\rm w} &= \sqrt{2 \nu_{\alpha\vert\rm w} {\rm \Delta} t_{|\rm w}} \mathcal{X} \,, \label{eq:dmuw}\\
    {\rm \Delta} p^x_{\alpha| \rm w} &= p_{\alpha| \rm w} {\rm \Delta} \mu_{\alpha| \rm w} + q_\alpha  E_x  {\rm \Delta}t_{\rm w} \nonumber \\
        & - \left( \beta_{\rm w\vert\rm s} \epsilon_{\alpha| \rm w} + p^x_{\alpha|\rm w} \right) \frac{{\rm d} u_{\rm w\vert s}}{{\rm d} t_{\vert\rm s}}{\rm \Delta}t_{\rm w} \label{eq:dpxw} \,,
\end{align}
where $\mu_{\alpha |\rm w }$ denotes the pitch-angle cosine relative to the shock normal, $p_{\alpha |\rm w}^x \equiv \mu_{\alpha |\rm w } p_{\alpha |\rm w}$ the longitudinal momentum, and $\mathcal{X}\sim\mathcal N(0,1)$ represents white noise, whose role is to simulate pitch-angle scattering in the microturbulence. This numerical scheme is thus similar to that of a 1D electrostatic PIC code. At every time step, the momentum of each particle is advanced in $\mathcal{R}_{\rm w}$ through Eqs.~\eqref{eq:dmuw} and \eqref{eq:dpxw}, where the discretized electric field is interpolated at the particle position. Moving to the shock frame, the particles positions are evolved using their updated momenta, and the electrostatic field is computed by a standard Poisson solver from the grid-projected particle charges. A complete description of the numerical method will be provided elsewhere \citep{Vanthieghem_inprep}.

\begin{figure*}[t!]
	\begin{center}
		\includegraphics[width=1.\textwidth]{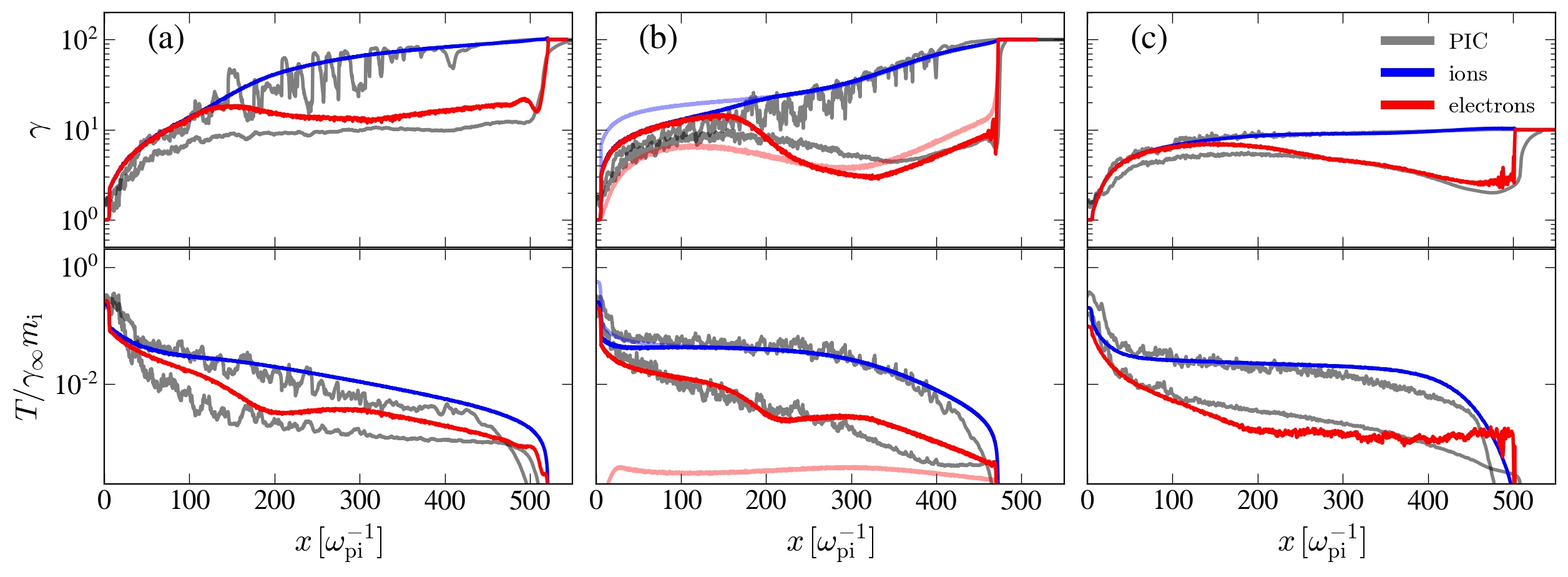}
		\caption{Comparison of the drift Lorentz factor (top) and temperature (bottom) between the PIC simulation with respective $(r,\gamma_\infty)\,=\,(25,100),\,(100,100),\,(100,10)$ (gray) and the MCP solution to the transport equation for the ions (blue) and electrons (red). The transport equation is integrated using the scattering frequency given by Eq.~\eqref{eq:MC_nu} with $\nu_{0\rm |w}\,\simeq\,6,\,6,\,0.2\,\epsilon_{\rm B}/k_\perp$. The light red and blue curves in the middle panel correspond to the solution to the transport equation neglecting the contribution of the electrostatic field; while they globally reproduce the observed amount of ion heating, they indicate insufficient electronic heating. We thus find that the shock dynamics is correctly reproduced for $\nu_{0\rm |w}\,\sim\,\epsilon_{\rm B}/k_\perp$ when accounting for the electrostatic field contribution. 
		}
		\label{fig:MC_nu}
	\end{center}
\end{figure*}

The above MCP model depends on the following ($x-$dependent) parameters: the scattering frequencies $\nu_{\alpha\vert\rm w}(x,p)$, the law of deceleration of the Weibel frame $u_{\rm w}(x)$ and the external charge $\rho_{\rm b}(x)$ imposed by the beam. Once these functional forms are fixed, the numerical integration gives the law of deceleration $u_{\rm e,i}(x)$ and heating $T_{\rm e,i}(x)$ of the electrons and ions of the background plasma, which can be compared with the outputs of the PIC simulations. 

The model equations are expressed in the shock frame, assuming stationarity, while our PIC simulations are time-dependent and run in the downstream rest frame. Given their long timescales, however, the precursor has reached, at least close to the shock, a near self-similar profile which depends only on $x - \beta_{\rm sh}t \propto x_{\vert\rm s}$. A fully consistent Lorentz transform from the downstream frame to the shock front frame would require the PIC plasma profiles to be recorded at multiple time steps. Here, we simply approximate the Lorentz boost by multiplying the space coordinate by $\gamma_{\rm sh } = \sqrt{4/3}$.

The scattering frequency $\nu_{\rm \vert w}$ in the Weibel frame is predicted to take on different forms depending on whether the particle is trapped or not in the magnetic filaments; known expressions  are summarized in Eqs.~(34) and (35) of \cite{Lemoine_2019_II}. For a unified description, our MCP model uses the following modified continuous formulation that retains the momentum dependencies in the trapped and untrapped limits:
\begin{equation}\label{eq:MC_nu}
  \setlength{\arraycolsep}{0pt}
  \nu_{\rm |w}(p_{\rm \vert w}) = \left\{ \begin{array}{l}
    \,\nu_{0\rm |w}\,\beta_{\rm |w} \left( p_{\rm |w}/m_{\rm i} \right)^{-2} ,\, \text{if} \quad p_{\rm |w} \geq p_{0\rm |w} \\
    \, \nu_{\vert \rm w}(p_{0\rm |w}) \left( \gamma_{\vert \rm w} /\gamma_{0 \vert \rm w} \right)^{-1} ,\, \text{else}
  \end{array} \right.
\end{equation}
with $p_{\vert \rm w}$ the particle momentum in the Weibel frame ($\gamma_{\vert\rm w}$ its Lorentz factor),
$\nu_{0\rm | w} \propto  \epsilon_{\rm B}/k_\perp$ a reference scattering frequency expressed in terms of $\epsilon_B = \langle B^2\rangle_y /4\pi m_{\rm i} n_{\rm i} \gamma_\infty^2$ the magnetic energy density fraction, and $k_\perp$ the dominant wavenumber in the transverse direction, {\it i.e.} the inverse of the filament size.
Particles transit from trapped to untrapped populations when their Larmor radius becomes comparable with the typical size of a filament, \emph{i.e.}, $p_{0\vert \rm w}/m_{\rm i} \simeq \epsilon_{\rm B}^{1/2} \omega_{\rm pi}/k_\perp$. As further detailed in \S\ref{App}, to solve this transport model, we fix the constant of proportionality between $\nu_{0\rm | w}$ and $\epsilon_{\rm B}/k_\perp$; we also extract the profile of $u_{\rm w}$ and $\rho_{\rm b}$, which we use directly in the MC-Poisson solver. 

We now discuss the results of this procedure, first comparing the scattering frequencies, as reconstructed along the history of the background plasma in the MCP model, with those extracted from the PIC simulation. For the latter, we apply known formulae [Eqs.~(34) and (35) of \cite{Lemoine_2019_II}]. The comparison is shown in Fig.~\ref{fig:nu}, which confirms that the reconstruction provides a satisfactory match to the scattering frequencies estimated from the PIC simulations. As one moves closer to the shock front, $\nu_{\rm e \vert w}$ decreases while $\nu_{\rm i \vert w}$ increases, by an order of magnitude or more, as a result of the evolution of the CFI spectrum, from electron kinetic scales at the tip of the precursor to ion kinetic scales near the shock front, in conjunction with strong electron heating;  $\nu_{\rm e \vert w}$ and $\nu_{\rm i \vert w}$ indeed meet near the shock front where the electrons reach near equipartition. 

Our model indeed reproduces the large amount of electron heating observed at the shock, for our three different simulations; this is illustrated in  Fig.~\ref{fig:MC_nu}. More precisely, for a scaling $\nu_{0|\rm w} \,\sim\, \epsilon_{\rm B}/k_\perp$ (meaning, with a prefactor not far from unity), it captures  the spatial profiles of the temperature and four-velocity for both ions and electrons of the background plasma. This satisfactory reconstruction holds across the whole precursor, including the abrupt electron slowdown at its tip, down to the fast dynamics inside the sub-shock layer. This finding represents the main result of our work. 

This numerical integration also reproduces fairly well the profile of the electric potential across the precursor, see the dashed lines in Fig.~\ref{fig:rho_phi}(a). Near the shock, where most of electron heating occurs, the reconstructed values lie within a factor $\sim2$ of the observed ones, within the shaded bands of variation.

That the ``best-fit'' values tying $\nu_{0|\rm w}$ to $\epsilon_{\rm B}/k_\perp$ in our numerical MC-Poisson reconstructions lie within an order of magnitude of unity provides additional support to our model, since $\nu_{0|\rm w} \,\sim\, \epsilon_{\rm B}/k_\perp$ is expected on theoretical grounds. We note some discrepancy between the prefactors used for the $\gamma_\infty=100$ simulations ($6$ for both) and that used for $\gamma_\infty=10$ ($0.2$). We do not expect this prefactor to depend on $\gamma_\infty$. However, we note that this ``best-fit'' value is subject to some uncertainty; it depends in particular on the extracted profiles of $k_\perp$, $\epsilon_B$, $u_{\rm w}$ and $\rho_{\rm b}$ from the PIC simulations (as discussed in \S\ref{App}). We also note that the notion of ``best-fit'' itself is somewhat vague: acceptable, albeit less satisfactory reconstructions of $u_{\rm i,e}$ and $T_{\rm i,e}$ can be obtained for values of the prefactor differing by up to an order of magnitude, as illustrated in Fig.~\ref{fig:MC_nu_var} of \S\ref{App}.  

We find that ions and electrons are heated through different mechanisms. For ions, the dominant contribution comes from collisionless Joule heating driven by the effective gravity in the decelerating turbulence frame, or equivalently, by stochastic interactions with the perpendicular motional electric fields if seen in the simulation frame:
the light blue line in Fig.~\ref{fig:MC_nu} (middle panel) indeed shows that $T_{\rm i}$ and $u_{\rm i}$ can be reproduced fairly well even when ignoring longitudinal electric fields. This is not altogether surprising as the ions are the dominant carriers of inertia and, in the case of a pair shock, for which this coherent electric field vanishes \citep{Pelletier_2019}, heating proceeds in the same manner. On the other hand, the electric field appears to provide the dominant source of heating for electrons. This is here illustrated by the light red line, which shows the corresponding evolution for electrons in the absence of $\langle E_x\rangle_y$; in that case, $T_{\rm e}$ at the shock lies $\gtrsim 2$ orders of magnitude below that observed in PIC simulations.

\section{Discussion and conclusions} \label{sec:discussion}

Our results thus indicate that the bulk of electron heating, up to near equipartition at relativistic, unmagnetized shock waves, is associated with the self-consistent generation of a coherent (along the transverse direction), longitudinal electric field across the shock precursor.
This field mainly originates from the charge separation imposed by the differential dynamics of the background plasma ions and electrons in the Weibel microturbulence.
In our theoretical description, electron heating results from a collisionless Joule process imparted by the coherent electric field in conjunction with pitch-angle scattering off the microturbulence, as expressed in the $\mathcal R_{\rm w}$ frame in which this microturbulence can be seen as essentially magnetostatic. The effective gravity force, felt by background ions and electrons in the decelerating $\mathcal R_{\rm w}$ frame, provides an additional source of heating; while subdominant for electrons, it accounts for most of the heating and deceleration of the ions, and therefore for the shock-crossing conditions.

Our conclusions rest on the successful comparison between a numerical integration of the above model and large-scale PIC simulations of relativistic electron-ion shocks, which we have conducted for values of the shock Lorentz factor up to $100$ and for mass ratios up to $100$.
Those simulations show the progressive heating of the ions and electrons up to quasi-equipartition, and they reveal the existence of a longitudinal coherent electric field across the shock precursor.  
To test our model,
we have extracted the velocity profile of the microturbulence frame $\mathcal R_{\rm w}$ and the beam charge profile. Using those estimates together with a general law for the scattering of particles in the microturbulence, we have performed a numerical integration of our model, which 
describes the stochastic heating and deceleration of a plasma in a decelerating Weibel-type turbulence, in the presence of a longitudinal electric field, by means of a Monte Carlo-Poisson solver. 
This integration compares satisfactorily to the PIC simulations: in particular, it reproduce fairly well the evolution of the electric potential as a function of distance to the shock, as well as the profiles of deceleration and heating of ions and electrons across the precursor. This suggests that this model captures the main features of electron and ion heating at weakly magnetized, relativistic shock waves.

In conclusion, let us point out some similarities and differences with  previous theoretical ideas on electron heating. Our model shares some features with that of \cite{Gedalin_2008} which describes electron heating in a cross-shock potential; in our scenario, however, the potential is rather found to extend over the precursor and heating is mostly stochastic, not directly associated with a DC jump as in that reference.
Moreover, the electric field originates from the difference in inertia, not from the growth of the CFI, as proposed in~\citep{Gedalin_2012, Kumar_2015}, nor from the broadband nature of that instability and its oblique variants~\citep{Gedalin_2008, Plotnikov_2013, Kumar_2015}.
We also note that the geometric configuration of the shock precursor is key to the emergence of this longitudinal electric field; a symmetric counterstreaming configuration, with two interpenetrating plasmas sharing similar physical characteristics, as simulated by \cite{Kumar_2015}, could not observe its presence. Finally, the origin of that electric field also differs from that envisaged in \cite{Milosavljevic_2006a} and \cite{Naseri_2018}, which attribute heating to the inductive electric field generated by the disruption of Weibel filaments.


\begin{acknowledgments}
AV thanks Frederico Fiuza and the anonymous referee for insightful suggestions and acknowledges support by the U.S. DOE Early Career Research Program under FWP 100331. The authors acknowledge financial support from the ILP Labex (reference ANR-10-LABX-63) as part of the Idex SUPER (reference ANR-11-IDEX-0004-02); the DIWINE Emergence SU 2019 programme;  the ANR-14-CE33-0019 MACH project; the ANR-20-CE30-0030 UnRIP project; the NSF grants PHY-1804048 and AST-1814708. Simulations were run on the HPC resources of TGCC/CCRT under the under the allocations 2018-A0030407666 and 2019-A0030407666 made by GENCI and on CORI at the National Energy Research Scientific Computing Center (NERSC) through ALCC award.
\end{acknowledgments}

%






\appendix

\section{Estimating the scattering frequency}\label{App}
\begin{figure}[ht]
	\begin{center}
		\includegraphics[width=0.95\textwidth]{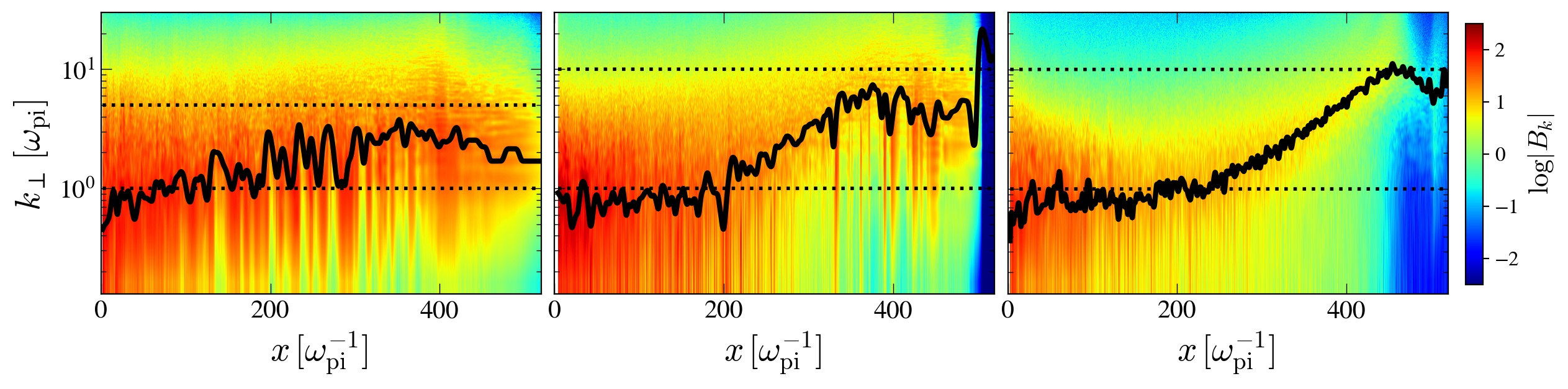}
		\caption{Longitudinal profile of the transverse Fourier modes of the out-of-plane ($B_z$) magnetic field across the shock precursor for $(r,\gamma_\infty)=(25,100),\,(100,100),\,(100,10)$ (from left to right). The two horizontal dotted lines correspond to the electron (top) and ion (bottom) plasma scales. The black line follows the dominant transverse wavenumber across the precursor, which is used to estimate the scattering frequency $\nu_{0|\rm w}$.
		}
		\label{fig:kperp}
	\end{center}
\end{figure}

\begin{figure}[ht]
\subfigure{\includegraphics[width=0.6\textwidth]{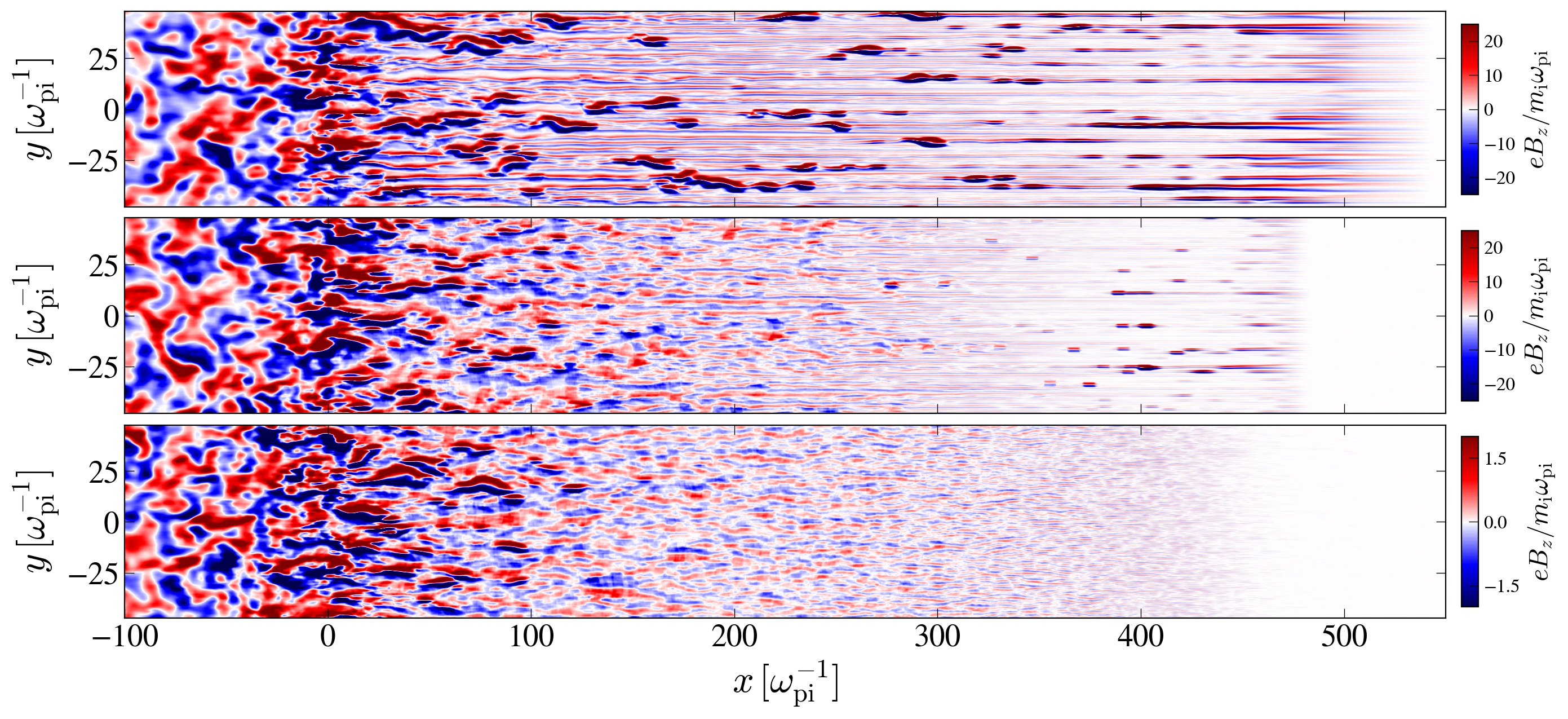}}
\subfigure{\raisebox{5mm}{\includegraphics[width=0.35\textwidth]{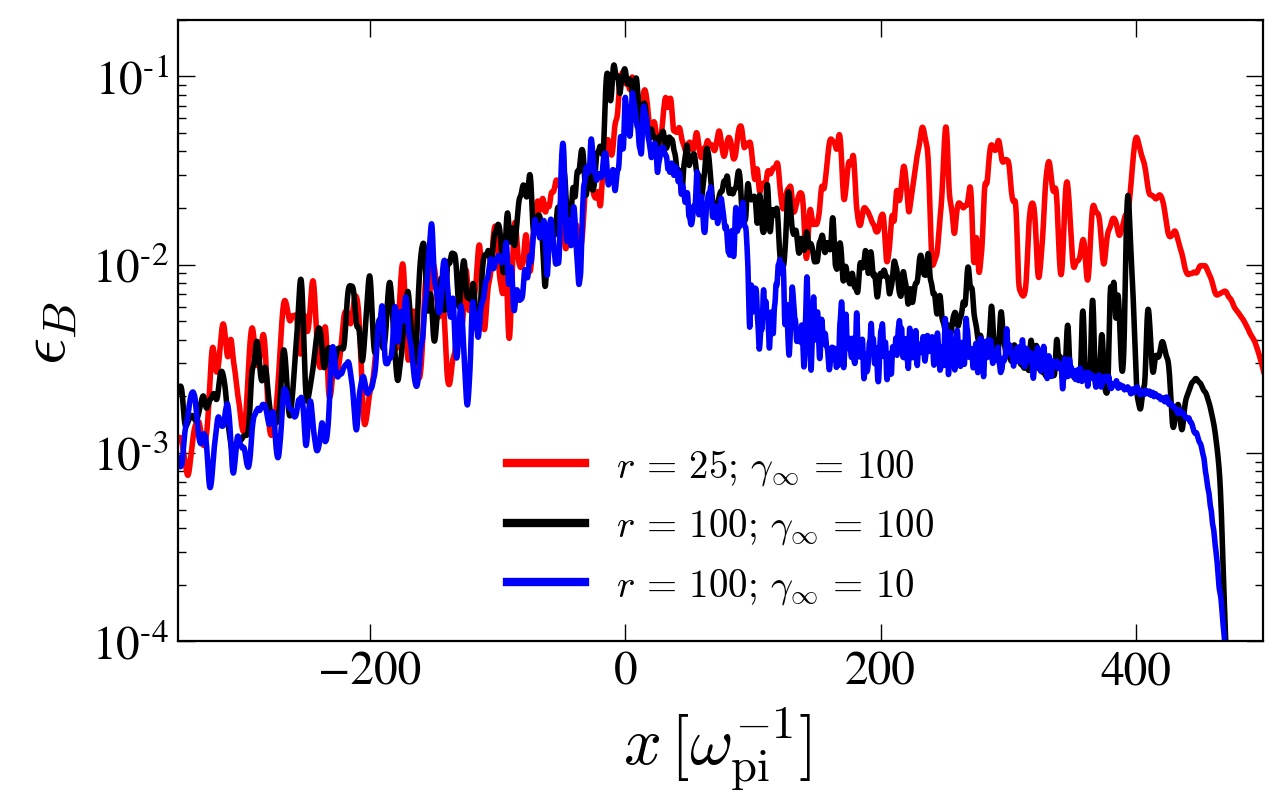}}}
\caption{Left: out-of-plane magnetic field component ($B_z$) in the simulation plane, for $(r,\gamma_\infty)=(25,100),\,(100,100),\,(100,10)$ (from top to bottom). Right: one-dimensional profile along the shock normal of the magnetic energy fraction $\epsilon_{\rm B}$, for the three simulations. This profile is used in the estimate of $\nu_{0|\rm w}$.}
\label{fig:Bz_eps}
\end{figure}

Here, we specify how the profiles of the physical quantities ($k_\perp$,  $\epsilon_{\rm B}$, $u_{\rm w}$, $\rho_{\rm b}$) used in our MC-Poisson model are extracted from the PIC simulations.
All other quantities then follow from the numerical integration. The dominant wavevector $k_\perp$ of the microturbulence is obtained from the Fourier space decomposition along the $y$-axis across the shock precursor, as illustrated in Fig.~\ref{fig:kperp}. One observes a progressive transition from electron skin depth scales at the tip of the precursor to ion skin depth scale close to the shock transition. The magnetization fraction $\epsilon_{\rm B}$ is obtained from the $y$-averaged magnetic field energy density. The magnetic field profiles for the different simulations are shown on the left of Fig.~\ref{fig:Bz_eps} while the associated magnetization level is shown on the right. Those quantities then serve to estimate the leading dependency of the scattering frequency, namely $\nu_{0\vert\rm w} \propto \epsilon_{\rm B}/k_\perp$, see Eq.~\eqref{eq:MC_nu}. The remaining prefactor, which we expect to be of the order of unity, is adjusted to reproduce the simulation results.

\begin{figure}[ht]
	\begin{center}
		\includegraphics[width=1\textwidth]{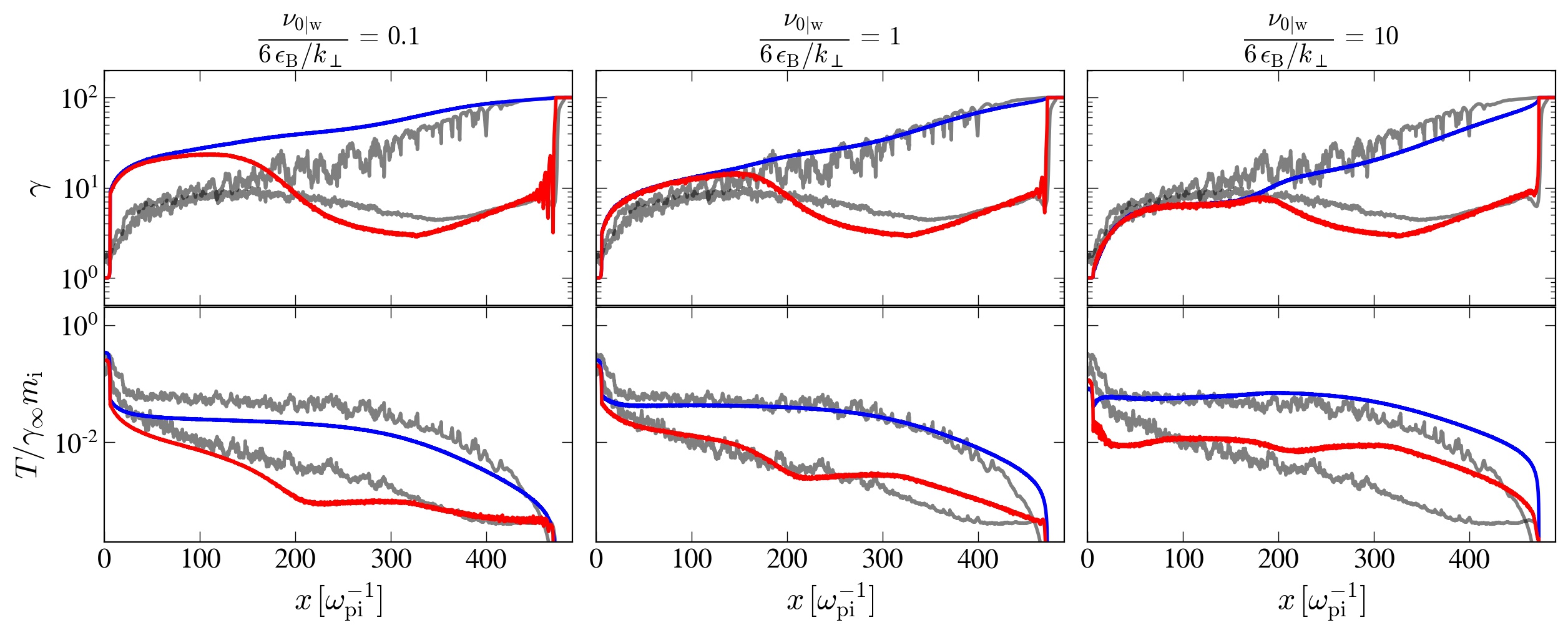}
		\caption{Comparison of the drift Lorentz factor (top) and temperature (bottom) between the PIC simulation with $(r,\gamma_\infty)=(100,100)$ (gray) and the MC-Poisson solution to the transport equation (including the electric field contribution) for the ions (blue) and the electrons (red). The transport equation is integrated using the scattering frequency given by Eq.~\eqref{eq:MC_nu} with, from left to right, $\nu_{0\rm |w}=0.6,\,6,\,60\,\epsilon_{\rm B}/k_\perp$. 
		}
		\label{fig:MC_nu_var}
	\end{center}
\end{figure}

To illustrate the dependence of our results on this prefactor, we focus on the simulation with the largest Lorentz factor and largest mass ratio, $(r,\gamma_\infty)=(100,100)$. We plot in Fig.~\ref{fig:MC_nu_var} the reconstructed history of the plasma four-velocity and temperature, as in Fig.~\ref{fig:MC_nu}, for values of $\nu_{0\vert\rm w}$ alternatively larger and smaller by an order of magnitude. As this prefactor departs from the value ($6$) used in the main text, the profiles start to differ from those measured in the PIC simulation, although the global amount of electron heating remains of the correct order of magnitude. This indicates that our results are not strongly sensitive to the choice of $\nu_{0\vert\rm w}$.

\bibliographystyle{mnras}
\bibliography{Bib}




\end{document}